\documentclass[aps,prl,twocolumn,superscriptaddress]{revtex4-1}

\usepackage{graphicx}
\usepackage{subfigure}
\usepackage[colorlinks=true,linkcolor=blue]{hyperref}%
\usepackage{color}

\begin{document}


\title{Structure of marginally jammed polydisperse packings of frictionless spheres}


\author{Chi Zhang}
\affiliation{Department of Physics, University of Fribourg, CH-1700 Fribourg, Switzerland}

\author{Cathal B. O'Donovan}
\affiliation{School of Physics, Trinity College Dublin, 2, Dublin, Ireland}

\author{Eric I. Corwin}
\affiliation{Department of Physics, University of Oregon, Eugene, Oregon 97403, USA}

\author{Fr\'ed\'eric Cardinaux}
\affiliation{Department of Physics, University of Fribourg, CH-1700 Fribourg, Switzerland}

\author{Thomas G. Mason}
\affiliation{Department of Chemistry and Biochemistry, University of California, Los Angeles, California 90095}
\affiliation{Department of Physics and Astronomy, University of California, Los Angeles, California 90095, USA.}

\author{Matthias E. M\"obius}
\affiliation{School of Physics, Trinity College Dublin, 2, Dublin, Ireland}

\author{Frank Scheffold}
\email[]{Frank.Scheffold@unifr.ch}
\affiliation{Department of Physics, University of Fribourg, CH-1700 Fribourg, Switzerland}


\date{\today}

\begin{abstract}We model the packing structure of a marginally jammed bulk ensemble of polydisperse spheres using an extended granocentric mode explicitly taking into account rattlers. This leads to a relationship between the characteristic parameters of the packing, such as the mean number of neighbors and the fraction of rattlers, and the radial distribution function $g(r)$. We find excellent agreement between the model predictions for $g(r)$ and packing simulations as well as experiments on jammed emulsion droplets. The observed quantitative agreement opens the path towards a full structural characterization of jammed particle systems for imaging and scattering experiments.
\end{abstract}




\maketitle
\begin{figure}
  \includegraphics[width=230pt]{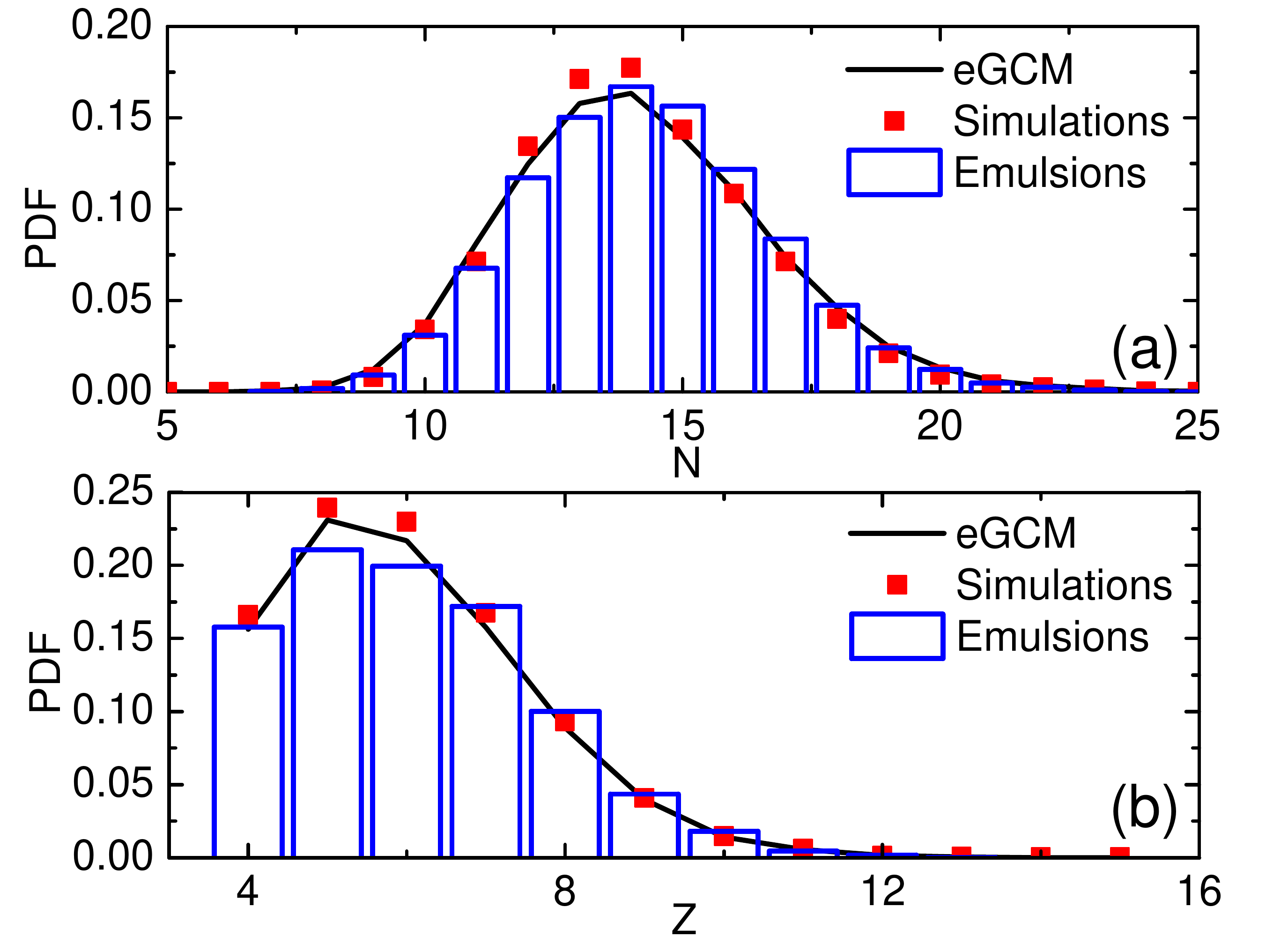}
  \caption{Probability distribution function for the particle neighbors (a) and contacts (b) for a polydispersity  PD=0.15. Solid line: eGCM; squares: simulations; bars: experiments on emulsion droplets. The eGCM and the simulations assume a lognormal size distribution. The size distribution of the emulsion is also close to lognormal with  $\text{PD} \simeq 0.15 \pm0.01$.}\label{FIGAR}
\end{figure}
\begin{figure}
  \includegraphics[width=260pt]{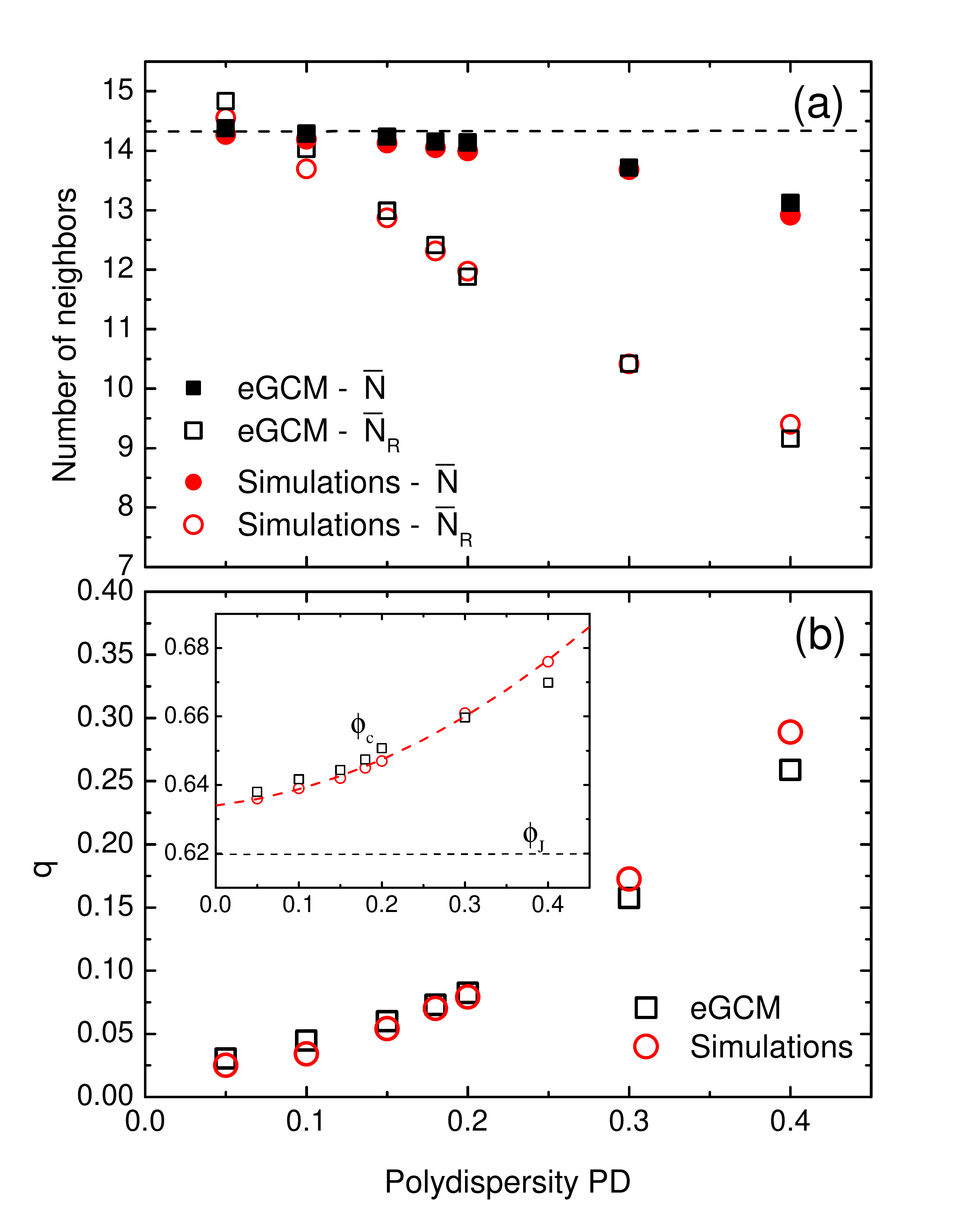}
  \caption{(a) Average number of first-shell neighbors $\bar N$ of jammed particles (full symbols) and rattlers (open symbols) versus polydispersity  (PD). Squares denote the results from the extended granocentric model (eGCM) and circles shows the data from numerical simulations. (b) Fraction of rattlers $q$ predicted by the eGCM (open squares) and from simulations (open circles). Inset: Predicted jamming volume fraction $\phi_c$ for all particles as a function of polydispersity. Dashed line: parabolic fit $\phi_c =0.634+0.0278 \cdot PD + 0.196 \cdot PD^2$.}  \label{FIG2R}
\end{figure}

The question how to optimally pack objects of various shape in space has been of fundamental interest in mathematics and physics for centuries \cite{torquato2001multiplicity,glotzer2007anisotropy}. It is also highly relevant for many practical problems ranging from storage and industrial packing to the properties of soft materials such as emulsions, foams or granular materials \cite{berthier2011dynamical,lacasse1996model, mason1997osmotic,hutzler2014z,tadros2009emulsion,stevenson2012foam,weaire1999physics,coussot2005rheometry,
mohan2012short}. Amorphous packings are particularly difficult to understand due to the complexity when dealing with disordered non-equilibrium structures. What has been known for a while is that disordered assemblies of spheres can be driven into a solid state by filling space up to a certain critical  volume fraction $\phi_c \sim 0.64$ \cite{bernal1959geometrical,scott1960packing}. At this point, denoted \emph{random close packing} or \emph{jamming}, the system is marginally stable. Mechanical stability is provided by an average isostatic number of contacts which in three dimensions is $\bar Z_J=6$ for frictionless spheres \cite{lacasse1996model,liu2010jamming,van2010jamming}. The advent of powerful simulation techniques over the last two decades has led to numerous new results and predictions. Soft spheres can be quenched into a compressed state $\phi>\phi_c$ and many relevant physical quantities, such as the modulus or the pressure, have been predicted to scale with the excess number of contacts $\Delta Z = \bar Z_J-6$ \cite{lacasse1996model,liu2010jamming}. Despite the recent progress made, the experimental relevance of these predictions has been questioned \cite{van2010jamming}. As matter of fact there are only few experimental studies which attempt to verify the numerical predictions \cite{majmudar2007jamming,jorjadze2013microscopic,zhang2009thermal} and study their relevance with respect to bulk properties of practically relevant materials \cite{scheffold2013linear,scheffold2014jamming}. This is partly due to the fact that the model assumptions made, such as the interaction potential between spheres or the size distribution, do not realistically reflect the situation in experimental systems such as in emulsions, dispersions or foams \cite{lacasse1996model, mason1997osmotic,hutzler2014z}.  Another important shortcoming of numerical studies is that these are generally carried out on real space assemblies whereas many experiments rely on scattering techniques that operate in $k-$space. While real-space experiments are rather straightforward in two dimensions \cite{majmudar2007jamming}, they are much more difficult to carry out in three dimensions \cite{jorjadze2013microscopic}, in particular when other physical properties, such as mechanical strength or internal dynamic modes, need to be studied as well. For this case scattering techniques, often in combination with mechanical shear measurements, have long been methods of choice to study soft disordered materials in the bulk \cite{lindner1991neutron,scheffold2009scattering,reufer2009temperature,stieger2004thermoresponsive}. Although scattering methods are highly appropriate for soft systems, dynamic methods can be highly sensitive to the rattlers in polydisperse systems, e.g. leading to ensemble-average mean square displacements that appear to relax more rapidly than is reflected by macroscopic rheology via a generalized Stoke-Einstein Relation (GSER)\cite{mason1995optical}. A direct comparison between numerical results and both structural as well as dynamic experiments however is again complicated by the idealisations made in the models. It would therefore be desirable to derive more general concepts that allow a direct comparison between experiment and theory. \newline \indent In the present work we address this problem and demonstrate how to model polydisperse sphere packings, taking into account explicitly the population of mechanically unstable particles, or \emph{rattlers}. To this end we expand on the granocentric model (GCM) introduced by Brujic and coworkers \cite{Brujic2009,newhall2011statistical} by statistically distributing non-contact neighbors and by taking into account size correlations between particle and shell. We show that such an extended granocentric model (eGCM) provides an accurate description of the statistical ensemble which in turn allows the comparison to  \emph{measureable} bulk quantities such as the radial distribution function $g(r)$. The latter is one of the most important structural measures for amorphous solids that is readily observable both in a real-space and in a scattering experiment, via the structure factor $S(k) = 1 + {{4\pi \rho } \mathord{\left/
 {\vphantom {{4\pi \rho } k}} \right.
 \kern-\nulldelimiterspace} k}\int_0^\infty  {dr\;r\sin \left( {kr} \right)\left[ {{g}\left( r \right) - 1} \right]} $, where $\rho$ is the particle number density \cite{hansen1990theory}. The importance of this quantity has recently been pointed out when studying the vestige of the jamming transition in an experiment both in two \cite{zhang2009thermal} and three dimensions  \cite{Thomas2013PRE} .
\newline \indent The granocentric model addresses the packing problem from the perspective of a single particle. The local packing structure is generated by the random formation of nearest neighbors \cite{Brujic2009}.  By numerical evaluation of the model predictions are made about the microscopic distributions of nearest neighbors and contacts. Existing variants of the model do not take into account rattlers \cite{Brujic2009,newhall2011statistical}. For our eGCM we divide the particles into two groups: mechanically stable jammed particles and freely floating rattlers, then take averages over a representative set of all  particles rather than considering only particles in contact. We first consider the the probability of finding a central particle with radius $a$ which is equal to particle size distribution $p(a)$. The polydispersity  PD = $\delta$a$/\bar a$ is defined by the standard deviation  $\delta a$ of $p(a)$ divided by its mean $\bar a=\int a \kern 1pt p(a) da$. For the probability to find a neighboring particle of a certain size $p_2(a)$ we explicitly consider the influence of size correlations between particles and their shell, previously neglected \cite{Brujic2009}. Packing simulations suggest \cite{Moebius13} $p_2(a)\propto p(a) \left[ {1+ (a/\bar a)^{2}} \right]$, which we are using here. Next we address the distribution $G_{s}(s)$ of surface-to-surface separations $s=r-2a$. The latter has to be modeled independently for the stable particles and for the rattlers. We can use the scaling of the excess number of contacts $\Delta Z \sim \sqrt{\bar Z_J-6}$ to derive  $G_{sJ}(s)\sim s^{-1/2}$(see also supplementary material) \cite{Wyart05An}. For the rattlers we assume their neighbors to be uniformly distributed $G_{sR}=const$.  The shell of neighbors is bounded by a cutoff distance $s_{\text{cutoff}}$. Here we denote with 'J' particles in contact, with 'R' rattlers and without suffix all particles.
 \newline \indent The eGCM can be evaluated numerically and we can obtain statistical information directly from the particle coordinates as shown below. However, a more general approach would be to reduce the discussion, e.g. of the radial distribution function, to its dependence on a small set of characteristic parameters, such as the average number of neighbors ($\bar N_J,\bar N_R$) of jammed or rattling particles, the fraction of rattlers $q$ and the distance $s_{\text{cutoff}}$. To this end we can write

\begin{equation} g(r)= \frac{1}{4\pi r^{2}\rho}\int_{} f(x)G_{s}(r-x)dx\label{Eqgvr}\end{equation}
were $f(x)=\int_{} p(a)p_2(x-a)da$ is the probability of finding a central particle with radius $a$ and another particle with radius $x-a$, for all possible $a$ and $G_{s}(s)=(1-q)G_{sJ}(s)+qG_{sR}(s)$. The link to the average number of neighbors is  established by normalization via $G_{sR}(s)=\bar N_{R}/s_{\text{cutoff}}$ and \begin{eqnarray}
\label{pdeltamain}
G_{sJ}(s)=
\left \{
\begin{array}{lll}
 6\delta(0) & (s=0), \\
 \frac{1}{2}(\bar N_{J}-6)(s_{\text{cutoff}} s)^{-\frac{1}{2}} & (s>0),
\end{array}
\right.
\end{eqnarray}
\newline \indent For a numerically evaluation of the extended granocentric model (eGCM) a limited set of input parameters is taken from packing simulations \cite{Moebius13} (see also supplementary material). The latter suggest isostatic values $\bar N_J=14.3$  and $\phi_J = 0.62$ for the jamming volume fraction, independent of polydispersity \cite{Moebius13,morse2014geometric}. Moreover we assume that the probability of contact for a particle added to the shell depends on the accessible solid angle $\Omega_{a}$. We find a best fit with simulations for a contact probability $ POC = \alpha \times (\Omega_{a}/\Omega_{max})^{3/2}$. Shells are filled up to a maximum solid angle $\Omega_{max}$ treated as an adjustable parameter in the numerical evaluation of the model. We generate 50000 neighboring shells following the generic approach of refs.\cite{Brujic2009,Moebius13}. Central particles with a contact number $Z<4$ are considered as rattlers. $\bar Z_J$ is obtained by taking the average over all central particles with $Z \ge 4$.  We adjust $\alpha,s_{\text{cutoff}},\Omega_{max}$ until $\bar Z_J, \bar N_J,\phi_J$ converge towards their isostatic values and thus obtain predictions for $N,Z,q,\phi_c$.  Moreover we find $s_{\text{cutoff}}/ \bar a \in \left[ {0.75,1} \right]$  and $\Omega_{max} \in \left[ {3.33\pi,3.53\pi} \right]$ (see also \cite{Brujic2009}).  As an example for the predictions of the model we plot in Figure \ref{FIGAR}  the  distribution of the number of neighbors $N$ and contacts $Z$ for a polydispersity  (PD) of $15\%$. In Figure \ref{FIG2R} the average values $\bar N$, $\bar N_{R}$ and $q$, are displayed as a function of polydispersity. We note immediately that $\bar N_{R}$ rapidly decreases with polydispersity.  This can be explained by the fact that with increasing polydispersity more rattling configurations are created by placing large particles next to a central particle. Since large particles occupy more solid angle, fewer neighbors can be placed around a rattler and thus $\bar N_{R}$ decreases.

\indent We first compare the eGCM-predictions to packing simulations of disordered packings of spheres. Details of the simulations are discussed elsewhere \cite{Moebius13}. Briefly, spheres are placed at random in a three-dimensional periodic cell and the size of the spheres is drawn from the distribution $p(a)$. The sphere sizes are then increased in unison until the desired packing fraction $\phi_c$ is reached.  Spheres are assumed to interact through purely repulsive body-centered forces and the overlap between two particles in contact leads to a harmonic interaction potential.  A conjugate gradient method is used to minimize the overlap between spheres and hence the total energy of the packing \cite{NumericalRecipesinC}.
 A comparison between the eGCM-predictions and packing simulations shows quantitative agreement, Figure \ref{FIGAR}. Equally good agreement is obtained for the average number of neighbors $\bar N, \bar N_{R}$, the fraction of rattlers and the jamming volume fraction, Figure \ref{FIG2R} and inset.
 \begin{figure}
  \includegraphics[width=210pt]{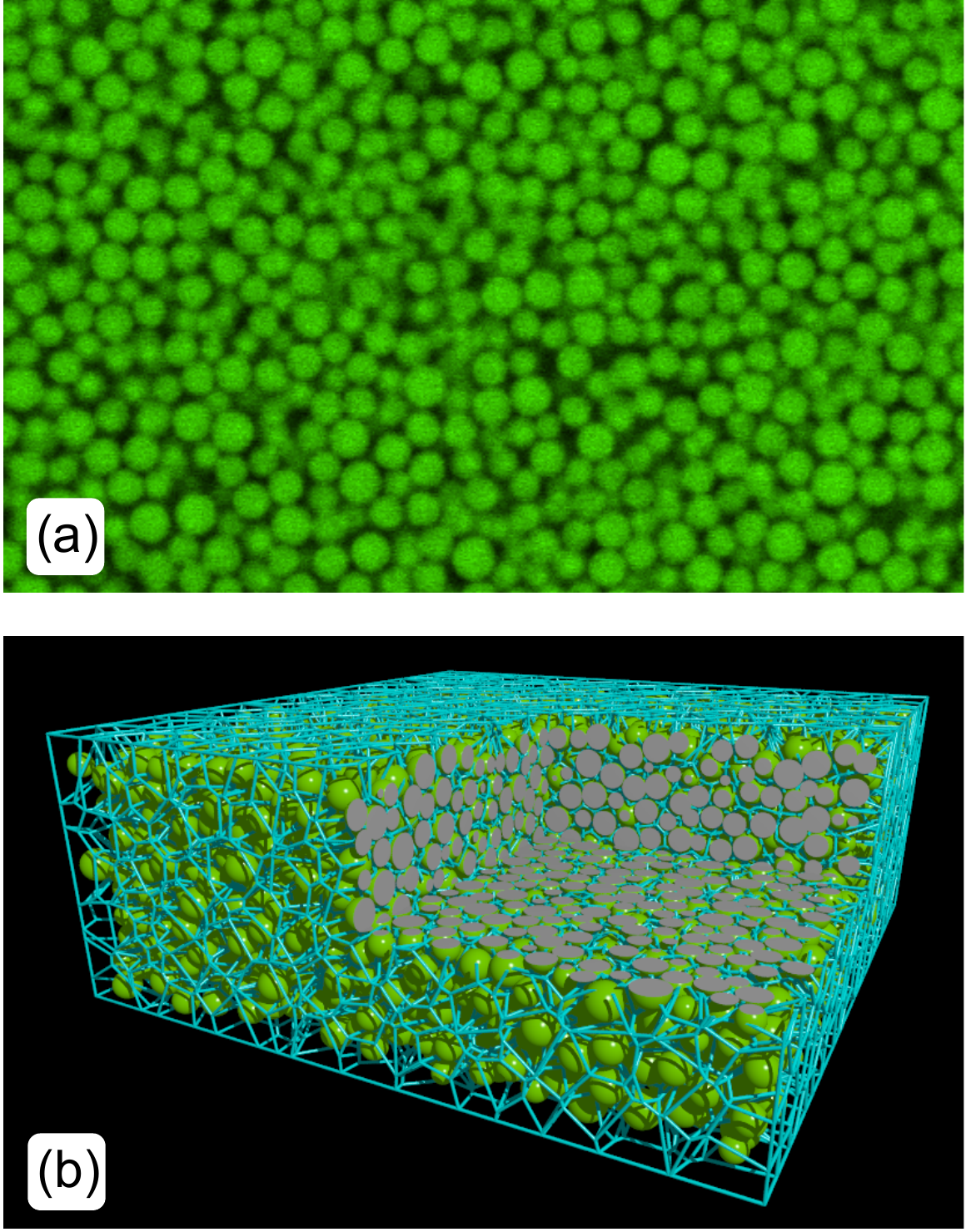}
  \caption{Three-dimensional imaging of jammed emulsions droplets. (a) Raw image of a plane in the bulk of the sample obtained by laser scanning confocal microscopy of light emitted by the fluorescent dye Nile-red at $\lambda=595$nm.  The droplets are marginally jammed and the volume fraction is $\phi \simeq 0.646 \pm 0.014$. (b) Three-dimensional reconstruction of the droplet sizes and positions using the sphere matching method (SMM). The lines show the Voronoi radical tessellation around the droplet centroids. The total dimensions are $51.2\mu $m$ \times 51.2 \mu $m$ \times 20.1\mu $m. One corner is cut out to reveal the internal structure of the jammed system}\label{FigExperiment}
\end{figure}
Next we compare the model predictions with experiments on micron scale emulsion droplets under marginal jamming conditions. For the experiments we prepare a 3:1 mixture by weight of PDMS and silicone oil (AR200) and emulsify it with sodium dodecyl sulfate (SDS) surfactant in water by shearing in a custom made Couette shear cell. Stabilized with SDS, the droplets are fractionated by size using depletion sedimentation \cite{Bibette91}. The size segregation is repeated until the desired polydispersity is reached. Subsequently the surfactant SDS is exchanged by the block-copolymer surfactant Pluronic F108, in order to sterically stablize the droplets. Finally Formamide and Dimethylacetamid (DMAC) are added to the solvent in order to match the density and refractive index simultaneously under experimental conditions at room temperature $T=22^\circ$C. Optical contrast between the droplets and the dispersion medium is obtained by adding the fluorescent dye Nile-red. Although the dye is present both in the solvent and the oil the emission spectra are different which allows to clearly distinguish both phases as shown in Figure \ref{FigExperiment} a.  The particle size and polydispersity are obtained from widefield microscopy. For the polydispersities considered we find the size distribution of the emulsion droplets to be close to log-normal. Equally, simulation data and the eGCM are evaluated for log-normal size distributions.  Here we include experimental data for three droplet radii  $ \bar a $ $=1.1\mu$m with a polydispersity PD=0.105,  $\bar a $ =$1.07\mu$m with a polydispersity PD=0.12 and $\bar a $ =$1.05\mu$m with a polydispersity PD=0.147 respectively. By lowering the temperature to 4$^\circ$C a slight density mismatch is induced and the sample can be spun down to densities at and above jamming. Several hundred microliters of the jammed sample are placed in a cylindrical cell tightly connected with UV-curable glue to a microscope cover slip which allows imaging from below in an inverted microscope.  High resolution images of the individual droplet positions are obtained using 3D laser scanning confocal microscopy (A1R, Nikon, Japan). The dye is excited with a $488$nm laser line and two emission channels ($525\pm50$nm and $595\pm50$nm) are recorded simultaneously to improve the quality of the analysis. 3D-images of size $512\times512\times201$ pixels are recorded with a resolution of 100nm/pixel in all spatial directions. For every stack of images, the acquisition time is $100$s. To track the position of the polydisperse droplets we implement the sphere matching methods (SMM) algorithm \cite{phdthesisBruijic2004}. A Voronoi radical tessellation is applied and particles with adjacent cell walls are identified as neighbors (Figure \ref{FigExperiment} b). We find the lateral position accuracy to be approximately $15$nm and axial accuracy $30$nm. In order to identify the point of marginal jamming the sample is diluted in steps of $\sim0.5\%$ in volume fraction. From a time series of 2D-images we can easily identify the liquid to solid transition, that in our case sets the jamming volume fraction $\phi_c$. From the droplet positions in 3D we calculate the radial distribution function  $g(r)$ and take an average over 20 image stacks in order to improve the statistical accuracy.

\begin{figure}
 \includegraphics[width=260pt]{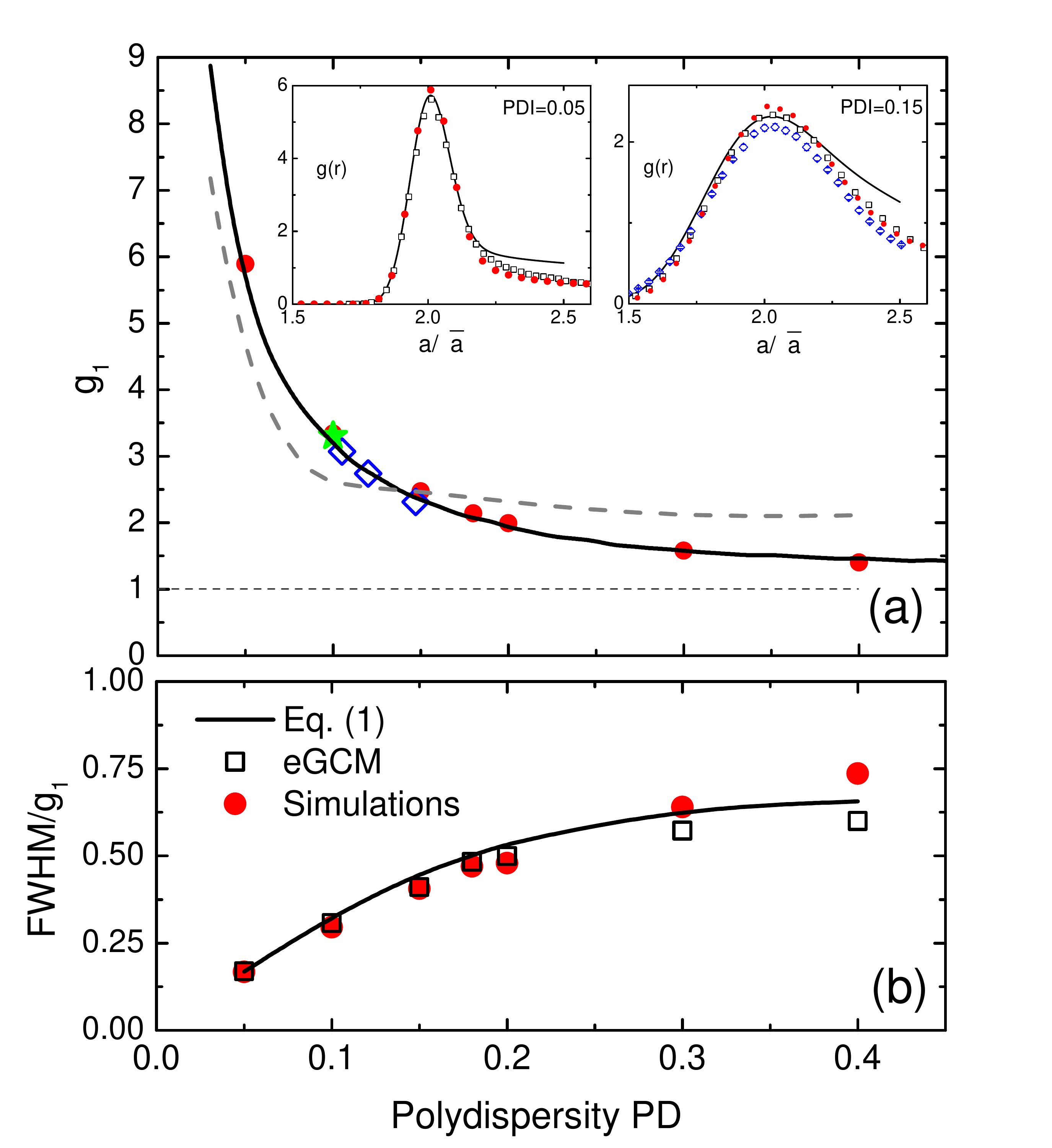}
 \caption{(a)  Polydispersity dependence of the first peak value $g_{1}$ of the radial distribution function. The solid line shows the prediction by the Eq.(\ref{Eqgvr}) and the dashed line the prediction of the original GCM \cite{Brujic2009}. Diamonds: experimental results for polydisperse emulsions. Full circles: packing simulations. Star: experimental results for microgel particles, ref. \cite{Thomas2013PRE}.  Inset: The first maximum of the radial distribution function $g(r)$ for a polydispersity  PD=0.05 (left) and PD=0.15 (right). Prediction by Eq.(\ref{Eqgvr}) (solid line), the eGCM (open squares), simulations (full circles) and the emulsion experiments (diamonds),  $\text{PD} \simeq 0.15 \pm0.01$ . (b) Normalized full-width half at maximum (FWHM).}\label{FIGgr2}
\end{figure}

In Figure \ref{FIGAR} we show the results obtained for the probability distribution of the number of neighbors $N$ and contacts $Z$ for a typical polydispersity of PD=0.15. The experimental results are in excellent agreement with both the packing simulations and the eGCM model. In Figure \ref{FIGgr2}(a) the model predictions for the radial distribution function are compared to the  $g(r)$ derived from the experimental droplet positions. While for perfectly monodisperse packings the peak value $g_1$ should diverge at the jamming transition this divergence is avoided for a size distribution of finite width.  We find quantitative agreement between all three data sets.  Extracting the peak value $g_1$ shows quantitative agreement  over a broad range of PD values as shown in  Figure \ref{FIGgr2}(a). Moreover we include a data point obtained for 3D assemblies of microgel particles with a mean size $\bar a \simeq 0.5\mu$m and a PD$\le$0.1 taken from \cite{Thomas2013PRE} and find again excellent agreement. Equally good agreement is obtained for the width of the first peak as shown in Figure \ref{FIGgr2}(b). Finally we note that in practice, for a known polydispersity,  $g(r)$ can be plotted directly using Eq.(\ref{Eqgvr},\ref{pdeltamain}) with input parameters $ \bar N_R, q$ taken from Figure \ref{FIG2R} and $s_{\text{cutoff}}/ \bar a\sim 0.8$.
\newline \indent The extensive comparison with experimental data and packing simulations demonstrates that the granocentric approach can deliver accurate predictions  for a bulk ensemble of marginally jammed particles covering the full range of polydispersities of practical interest. Quantitative modeling of $g(r)$ and also $S(q)$ in turn will allow a more direct comparison between experiments and the underlying structure of the packing. A future extension of the model towards higher densities, taking into account finite particle compression, would open the path towards a full structural characterization of the bulk ensemble of polydisperse jammed particle systems.
\newline \indent CZ and FS acknowledge financial support by the Swiss National Science Foundation (Project No. 149867). EIC acknowledges the support of the NSF under CAREER Award No. DMR-1255370. M.M. and C.O. acknowledge financial support from the Science Foundation Ireland, grant no. 11/RFP/MTR/3135

\clearpage

\end{document}